\documentclass[twocolumn,showpacs,amsmath,amssymb]{revtex4}
\usepackage{mathrsfs}
\usepackage{graphicx}
\usepackage{amsfonts}
\usepackage{dcolumn}
\usepackage{bm}
\usepackage{multirow}
\usepackage{CJK}

\begin{document}
\title{\textbf{Phase transitions in a multistate majority-vote model on complex networks}}

\author{Hanshuang Chen}\email{chenhshf@ahu.edu.cn}

\author{Guofeng Li}

\affiliation{School of Physics and Materials Science, Anhui
University, Hefei, 230601, China}

\date{\today}

\begin{abstract}
We generalize the original majority-vote (MV) model from two states
to arbitrary $p$ states and study the order-disorder phase
transitions in such a $p$-state MV model on complex networks. By
extensive Monte Carlo simulations and a mean-field theory, we show
that for $p\geq3$ the order of phase transition is essentially
different from a continuous second-order phase transition in the
original two-state MV model. Instead, for $p\geq3$ the model
displays a discontinuous first-order phase transition, which is
manifested by the appearance of the hysteresis phenomenon near the
phase transition. Within the hysteresis loop, the ordered phase and
disordered phase are coexisting and rare flips between the two
phases can be observed due to the finite-size fluctuation. Moreover,
we investigate the type of phase transition under a slightly
modified dynamics [Melo \emph{et al.} J. Stat. Mech. P11032 (2010)].
We find that the order of phase transition in the three-state MV
model depends on the degree heterogeneity of networks. For $p\geq4$,
both dynamics produce the first-order phase transitions.

\end{abstract}
 \maketitle

\section{Introduction}\label{sec1}
Spin models such as the Ising model play fundamental roles in
studying phase transitions and critical phenomena in the field of
statistical physics \cite{Baxter1989}. They have also significant
implications for understanding various social and biological
phenomena where co-ordination dynamics is observed, e.g., in
consensus formation and adoption of innovations
\cite{AJP08000470,RMP08001275,RMP09000591}. The spin orientations
can represent the choices made by an agent on the basis of
information about its local neighborhood. Along these lines, so much
has been done in recent years in social systems from human
cooperation \cite{PR.687.1,PLA.380.2803} to vaccination
\cite{wang2016statistical,RevModPhys.87.925} to crime
\cite{PLR.12.1} and saving human lives \cite{JSP.158.735}, as well
as biological systems from collective motion \cite{PR.517.71} to
transport phenomena \cite{Rep.Prog.Phys.74.116601} to criticality
and dynamical scaling \cite{2017arXiv171204499}.

The majority-vote (MV) model is one of the simplest nonequilibrium
generalizations of the Ising model \cite{JSP1992}. In the model,
each spin is assigned to a binary variable. At each time step, each
spin tends to align with the local neighborhood majority but with a
noise intensity $f$ giving the probability of misalignment. The MV
model not only plays an important role in the study of
nonequilibrium phase transitions, but it also help to understand
opinion dynamics in social systems \cite{RMP09000591}. The two-state
MV model has been extensively studied in various interacting
substrates, such as regular lattices
\cite{PhysRevE.75.061110,PhysRevE.81.011133,PhysRevE.89.052109,PhysRevE.86.041123,PhysRevE.95.012101},
random graphs \cite{PhysRevE.71.016123,PA2008}, small-world networks
\cite{PhysRevE.67.026104,IJMPC2007,PA2015}, scale-free networks
\cite{IJMPC2006(1),IJMPC2006(2),PhysRevE.91.022816}, modular
networks \cite{CPL2015}, complete graphs \cite{PhysRevE.96.012304},
and spatially embedded networks \cite{PhysRevE.93.052101}. With the
exception of an inertial effect that was considered
\cite{PhysRevE.95.042304,Chaos27.081102,PhysRevE.96.042305}, all the
previous studies have shown that the two-state MV model presents a
continuous second-order phase transition at a critical value of $f$.

The multistate MV model is a natural generalization of the two-state
case, As its equilibrium counterpart, the Potts model is a
generalization of the Ising model \cite{RevModPhys.54.235}. The
three-state MV model on a regular lattice was considered in
\cite{PhysRevE.60.3666,JPA2002}, where the authors found that the
critical exponents for this non-equilibrium model are in agreement
with the ones for the equilibrium three-state Potts model,
supporting the conjecture of \cite{PhysRevLett.55.2527}. Melo
\emph{et al.} studied the three-state MV model on random graphs and
showed that the phase transition is continuous and the critical
noise is an increasing function of the mean connectivity of the
graph \cite{JSM2010}. Li \emph{et al.} studied a three-state MV
model with a slightly different dynamics in an annealed random
network, and they showed the phase transition belongs to a
first-order type \cite{JSM2016.073403}. Lima introduced an
unoccupied state to the two-state MV model in square lattices and
found that this model also falls into the Ising universality
\cite{PA2012}. Costa \emph{et al.} generalized the state variable of
the MV model from a discrete case to a continuous one, and found
that a Kosterlitz-Thouless-like phase appears in low values of noise
\cite{PhysRevE.71.056124}.

In the present work, we generalize the MV model to arbitrary
multiple states, and we focus on the natures of phase transitions in
the multi-state MV model on complex networks. By Monte Carlo (MC)
simulation, we show that if the number of states is greater than or
equal to 3, a clear hysteresis loop is observed as noise is dialed
up and down, which is a typical feature of a first-order phase
transition. Moreover, we propose a mean-field theory to validate the
simulation results. Finally, we investigate the type of phase
transition under a slightly modified dynamics
\cite{PhysRevE.60.3666,JPA2002,JSM2010}. We find that such a small
difference in dynamics leads to the essential difference in the type
of phase transition in the three-state MV model on Erd\"os-R\'enyi
(ER) random networks or higher degree heterogeneous networks.

\section{Model}\label{sec2}
We generalize the original MV model from two states to arbitrary
multiple states. The model is defined on an unweighted network with
size $N$ described by an $N\times N$ adjacency matrix, whose
elements $A_{ij}=1$ if a directed edge is emanated from node $j$ and
ended at node $i$, and $A_{ij}=0$ otherwise. Each node $i$ can be in
any of the $p$ states: ${\sigma _i} \in \{ 1, \cdots ,p\}$. The
number of the neighbors of node $i$ in each state $\alpha$ can be
calculated as ${n_i^\alpha} = \sum\nolimits_{j =1}^N
A_{ji}{\delta(\sigma_j-\alpha)}$, where $\delta(x)$ is the Kronecker
symbol defined as $\delta=1$ if $x=0$ and $\delta=0$ otherwise.

In the following, we introduce two slightly different types of
dynamical rules. For both dynamics, the node $i$ take the same value
as the majority spin with the probability $1-f$, i.e., ${\sigma _i}
= {\left. \alpha \right|_{{n_i^\alpha } = \max \{ {n_i^1}, \cdots
,{n_i^q}\} }}$. With the supplementary probability $f$, the node $i$
takes the same value as the minority spin, i.e., ${\sigma_i} =
{\left. \alpha \right|_{{n_i^\alpha } = \min \{ {n_i^1}, \cdots
,{n_i^q}\} }}$ for type-I dynamics. For type-II dynamics, the node
$i$ takes the same value as that of nonmajority spins (not
necessarily the minority spin), i.e., ${\sigma_i} = {\left. \alpha
\right|_{{n_i^\alpha } \neq \max \{ {n_i^1}, \cdots ,{n_i^q}\} }}$.
If more than one candidate state is in the majority spin or in the
minority spin, we randomly choose one of them. Here, the probability
$f$ is called the noise intensity, which plays a similar role to the
temperature in equilibrium systems and measures the probability of
disagreeing with the majority of neighbors. For convenience, the
former and the latter are called the type-I and type-II
\emph{$p$-state MV model}, respectively. If $p=2$, both dynamics are
mutually equivalent and recover to the original two-state MV model.
We should note that the type-II three-state MV model shows
continuous phase transitions on square lattices
\cite{PhysRevE.60.3666,JPA2002} and ER random networks
\cite{JSM2010}.

To characterize the critical behavior of the model, we define the
order parameter as the modulus of the magnetization vector, that is,
$m = {\left( {\sum\nolimits_{\alpha  = 1}^p {m_\alpha ^2} }
\right)^{1/2}}$, whose components are given by
\begin{eqnarray}
{m_\alpha } = \sqrt {\frac{p}{{p - 1}}} \left[
{\frac{1}{N}\sum\limits_i {\delta \left( {\alpha  - {\sigma _i}}
\right) - \frac{1}{p}} } \right], \label{eq1}
\end{eqnarray}
where the factor $\sqrt{p/(p - 1)}$ is used to normalize the
magnetization vector.

\section{Results}\label{sec3}
\subsection{Type-I dynamics} 
\begin{figure}
\centerline{\includegraphics*[width=1.0\columnwidth]{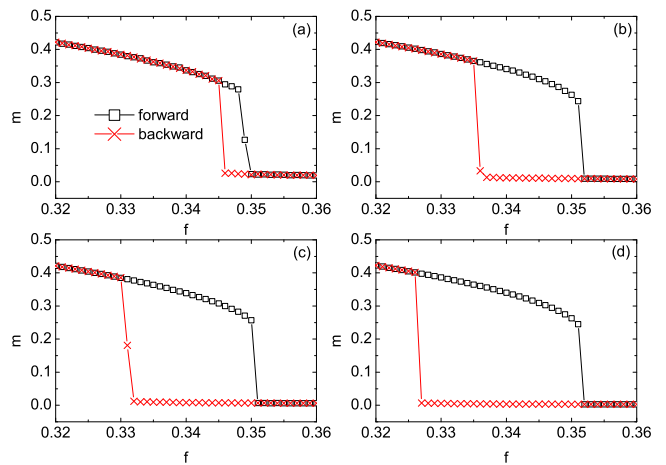}}
\caption{(Color online). First-order phase transition in the type-I
three-state MV model on ER networks, characterized by a hysteresis
loop of $m$ as noise intensity $f$ is dialed up and down. From
(a)-(d), the network sizes are $N=10^4$, $5 \times 10^4$, $10^5$,
and $5 \times 10^5$, respectively. The average degree is fixed at
$\left \langle k \right \rangle=10$. Squares ($\Box$) and crosses
($\times$) correspond to forward and backward simulations,
respectively. \label{fig1}}
\end{figure}

We first focus on the type-I three-state MV model. By performing
extensive MC simulations on ER random networks \cite{ER1960}, we
show the magnetization $m$ as a function of the noise intensity $f$,
as shown in Fig. \ref{fig1}. The network size $N$ varies from Fig.
\ref{fig1}(a) to Fig. \ref{fig1}(d): $N=10^4$ (a), $N=5 \times 10^4$
(b), $N=10^5$ (c), and $N=5 \times 10^5$ (d). The average degree
$\left \langle k \right \rangle=10$ is kept unchanged. The
simulation results are obtained by performing forward and backward
simulations, respectively. The former is done by calculating the
stationary value of $m$ as $f$ increases from 0.32 to 0.36 in steps
of 0.001 and using the final configuration of the last simulation
run as the initial condition of the next run, while the latter is
performed by decreasing $f$ from 0.36 to 0.32 with the same step.
One can see that as $f$ increases, $m$ abruptly jumps from nonzero
to zero at $f = f_{cF}$, which shows that a sharp transition takes
place for the order-disorder transition. Additionally, the curve
corresponding to the backward simulations also shows a sharp
transition from the disordered phase to the ordered phase at $f =
f_{cB}$. These two sharp transitions occur at two different values
of $f$, leading to a hysteresis loop with respect to the dependence
of $m$ on $f$. The hysteresis loop becomes clearer as the network
size increases. Such a feature indicates that a discontinuous
first-order phase transition occurs in the type-I three-state MV
model. This is in contrast to the original two-state MV model in
which a continuous second-order phase transition was observed
\cite{PhysRevE.71.016123,PA2008,PhysRevE.91.022816}.

\begin{figure}
\centerline{\includegraphics*[width=1.0\columnwidth]{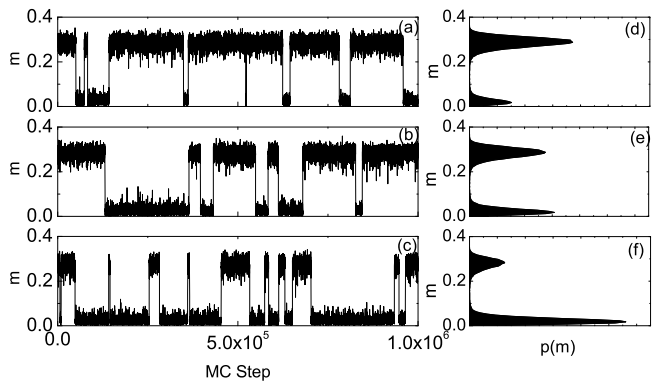}}
\caption{Coexistence of ordered and disordered phases in the
hysteresis region for the type-I three-state MV model. Three typical
time series of the magnetization $m$ on an ER network, corresponding
to three different noise intensities chosen from the hysteresis
region: $f=0.3472$ (a), $f=0.3475$ (b), and $f=0.3480$ (c). (d)-(f)
show the corresponding histograms for the distribution of $m$ at the
three values of $f$ as in (a-c), respectively. The networks
parameters are $N=10^4$ and $\left \langle k \right \rangle=10$.
\label{fig2}}
\end{figure}

To further verify the first-order nature of phase transition in the
type-I three-state MV model, in Fig. \ref{fig2}(a-c) we show three
long time series of $m$ corresponding to three distinct $f$ on an ER
network with $N=10^4$ and $\left \langle k \right \rangle=10$. Here
the noise intensity $f$ is chosen from the hysteresis region. One
can see that in the hysteresis region the ordered and disordered
phases are coexisting. Due the finite-size fluctuation, phase
flipping between the ordered phase and the disordered phase can be
rarely observed. As $f$ increases, the system spends more time on
the disordered phase. In Fig. \ref{fig2}(d-f), we show the
corresponding histograms for the distribution of $m$ at the three
distinct $f$ as in Fig. \ref{fig2}(a-c). All the distributions are
bimodal with a peak around $m=0$ and the other one at $m>0$. On the
other hand, with the increase of $f$ the peak around $m=0$ becomes
higher, indicating that the disordered phase becomes more stable
with $f$. In general, as $N$ increases the fluctuation level becomes
less significant and the mean time of phase switching increases
exponentially with $N$, so that it is difficult to observe the phase
flipping in the allowable computational time for larger $N$.

In \cite{PhysRevE.91.022816}, we developed a heterogeneous
mean-field theory to deal with the two-state MV model on degree
uncorrelated networks, and we derived that the critical noise is
determined by the ratio of the first-order moment to the 3/2-order
moment of degree distribution. Also for the two-state MV model, a
quenched mean-field theory was proposed recently
\cite{EPL.2017.120.18003}, which showed that the critical noise is
determined by the largest eigenvalue of a deformed network adjacency
matrix. In \cite{JSM2016.073403}, we proposed a simple mean-field
theory for the three-state MV model on a degree-regular random
network in which each node is randomly connected to exactly $k$
neighbors, and degree distribution follows the $\delta$-function. In
the following, we shall develop a heterogeneous mean-field theory
that is applicable not only for any number $p$ of the states, but
also for any degree distribution without degree-degree correlation.

To this end, let $x_k^\alpha$ denote the probability of nodes of
degree $k$ being in the state $\alpha$. The dynamical equation for
$x_k^\alpha$ reads,
\begin{eqnarray}
{\dot x_k^\alpha } = \sum\limits_{\beta  \ne \alpha } {{x_k^\beta }}
{w_k^{\beta  \to \alpha }} - {x_k^\alpha }\sum\limits_{\beta  \ne
\alpha } {{w_k^{\alpha  \to \beta }}}, \label{eq2}
\end{eqnarray}
where $w_k^{\alpha  \to \beta }$ is the transition probability of
nodes of degree $k$ from the state $\alpha$ to the state $\beta$.
According to the definition of the MV model, the probability
$w_k^{\alpha \to \beta }$ can be written as the sum of two parts,
\begin{eqnarray}
{w_k^{\alpha  \to \beta }} = (1 - f){P_k^\beta } + f{{\tilde
P}_k^\beta }, \label{eq3}
\end{eqnarray}
where the first part is given by the probability $1-f$ of nodes of
degree $k$ taking the majority-rule, multiplied by the probability
$P_k^\beta$ that the state $\beta$ is the majority state among the
neighbors of nodes of degree $k$. Likewise, the second part is the
product of the probability $f$ of nodes of degree $k$ taking the
minority-rule and the probability $\tilde P_k^\beta$ that the state
$\beta$ is the minority state. Utilizing the normalization
conditions, $\sum\nolimits_\beta {{x_k^\beta }}=\sum\nolimits_\beta
{{P_k^\beta }} = \sum\nolimits_\beta {{{\tilde P}_k^\beta }}  = 1$,
Eq. (\ref{eq2}) can be simplified to
\begin{eqnarray}
{{\dot x}_k^\alpha } =  - {x_k^\alpha } + (1 - f){ {P}_k^\alpha } +
f{{\tilde P}_k^\alpha }. \label{eq4}
\end{eqnarray}
In the steady state, $\dot x_k^\alpha=0$, we have
\begin{eqnarray}
{x_k^\alpha } = (1 - f){P_k^\alpha } + f{{\tilde P}_k^\alpha }.
\label{eq5}
\end{eqnarray}

Let us further define $X_\alpha$ as the probability that for any
node in the network, a randomly chosen nearest-neighbor node is in
the state $\alpha$. For degree uncorrelated networks, the
probability that a randomly chosen neighboring node has degree $k$
is $kP(k)/\left\langle k \right\rangle$ \cite{RMP08001275}, where
$P(k)$ is degree distribution defined as the probability that a node
chosen at random has degree $k$ and $\left\langle k
\right\rangle=\sum\nolimits_k {kP(k)}$ is the average degree.
Therefore, The probabilities $x_k^\alpha$ and $X_\alpha$ satisfy the
relation
\begin{eqnarray}
{X_\alpha } = \sum\limits_k {\frac{{kP(k)}}{{\left\langle k
\right\rangle }}} x_k^\alpha. \label{eq6}
\end{eqnarray}

Let $n_\alpha$ denote the number of neighbors of a node of degree
$k$ in the state $\alpha$, and the probability of a given
configuration $\{n_\alpha\}$ can be expressed as a multinominal
distribution,
\begin{eqnarray}
{\Xi^k _{{n_1}, \ldots ,{n_p}}}\left( {{X_1}, \ldots ,{X_p}} \right)
= \frac{{k!}}{{\prod\limits_\alpha  {{n_\alpha }!}
}}\prod\limits_\alpha  {X_\alpha ^{{n_\alpha }}}, \label{eq7}
\end{eqnarray}
where $k=\sum \nolimits_\alpha n_\alpha$. Therefore, $P_k^\alpha$
and $\tilde P_k^\alpha$ can be written as
\begin{eqnarray}
P_k^\alpha  = \sum\limits_{\left. {\left\{ {n_\alpha } \right\}}
\right|n_\alpha  \geq n_\beta ,\forall \beta  \ne \alpha }
{\frac{1}{{1 + \Omega_\alpha \left( {\left\{ {n_\alpha } \right\}}
\right)}} \Xi ^k_{{n_1}, \ldots ,{n_p}}  },\label{eq8}
\end{eqnarray}
and
\begin{eqnarray}
\tilde P_k^\alpha  = \sum\limits_{\left. {\left\{ {n_\alpha }
\right\}} \right|n_\alpha  \leq n_\beta ,\forall \beta  \ne \alpha }
{\frac{1}{{1 + \Omega_\alpha \left( {\left\{ {n_\alpha } \right\}}
\right)}} \Xi ^k_{{n_1}, \ldots ,{n_p}}  },\label{eq9}
\end{eqnarray}
where $\Omega_\alpha \left( {\left\{ {n_\alpha } \right\}} \right) =
\sum\nolimits_{\beta  \ne \alpha } {\delta \left( {{n_\beta } -
{n_\alpha }} \right)}$ is the number of states whose number of nodes
is the same as $n_\alpha$. If $\Omega_\alpha=0$, the state $\alpha$
is the only majority (minority) state, such that the factor
$1/(1+\Omega_\alpha)$ in Eq. (\ref{eq8}) (Eq. (\ref{eq9})) equals to
one. If $\Omega_\alpha=1$, there are two candidate majority
(minority) states, such that the factor is equal to $1/2$, and so
forth.

Substituting Eq. (\ref{eq5}) into Eq. (\ref{eq6}), we arrive at a
set of self-consistent equations of $X_\alpha$,
\begin{eqnarray}
{X_\alpha } = (1 - f)\sum\limits_k {\frac{{kP(k)}}{{\left\langle k
\right\rangle }}} P_k^\alpha  + f\sum\limits_k
{\frac{{kP(k)}}{{\left\langle k \right\rangle }}} \tilde P_k^\alpha.
\label{eq10}
\end{eqnarray}
Notice that $X_\alpha=1/p$ is always a set of solutions of Eq.
(\ref{eq10}) since $P_k^\alpha=\tilde P_k^\alpha=1/p$ at
$X_\alpha=1/p$. Such a trivial solution corresponds to the
disordered phase ($m=0$). For convenience, the trivial solution is
denoted by a vector
$\textbf{X}=\textbf{X}^*\equiv(1/p,\cdots,1/p)^\top$, where the
superscript $\top$ denotes the transpose. To evaluate the stability
of $\textbf{X}^*$, we need to write down the Jacobian matrix
$\textbf{J}$ of Eq. (\ref{eq10}). Since $X_\alpha$ satisfies the
normalization condition $\sum\nolimits_\alpha {X^\alpha}=1$, only
$p-1$ variables among $X_\alpha$ ($\alpha=1,\cdots,p$) are
independent of each other. To the end, we select
$X_1,\cdots,X_{p-1}$ as the independent variables and therefore
$\textbf{J}$ is a $(p-1)$ dimensional square. The matrix elements of
$\textbf{J}$ are given by
\begin{eqnarray}
{J_{\alpha \beta }} = \left( {1 - f} \right)\sum\limits_k
{\frac{{kP(k)}}{{\left\langle k \right\rangle }}} {\left.
{\frac{{\partial P_k^\alpha }}{{\partial {X_\beta }}}}
\right|_{{\textbf{X}^*}}} + f\sum\limits_k
{\frac{{kP(k)}}{{\left\langle k \right\rangle }}} {\left.
{\frac{{\partial \tilde P_k^\alpha }}{{\partial {X_\beta }}}}
\right|_{{\textbf{X}^*}}}, \nonumber \\ \label{eq11}
\end{eqnarray}
with $\alpha,\beta=1,\cdots,p-1$. According to Eq. (\ref{eq8}) and
Eq. (\ref{eq9}), we have
\begin{eqnarray}
{\left. {\frac{{\partial P_k^\alpha }}{{\partial {X_\beta }}}}
\right|_{{\textbf{X}^*}}} = \sum\limits_{\left. {\left\{ {{n_\alpha
}} \right\}} \right|{n_\alpha } \geq {n_\beta },\forall \beta \ne
\alpha } {\frac{1}{{1 + {\Omega _\alpha }\left( {\left\{ {{n_\alpha
}} \right\}} \right)}}} {\left. {\frac{{\partial \Xi _{{n_1}, \ldots
,{n_p}}^k}}{{\partial {X_\beta }}}}
\right|_{{\textbf{X}^*}}},\nonumber \\\label{eq12}
\end{eqnarray}
and
\begin{eqnarray}
{\left. {\frac{{\partial \tilde P_k^\alpha }}{{\partial {X_\beta
}}}} \right|_{{\textbf{X}^*}}} = \sum\limits_{\left. {\left\{
{{n_\alpha }} \right\}} \right|{n_\alpha } \leq {n_\beta },\forall
\beta \ne \alpha } {\frac{1}{{1 + {\Omega _\alpha }\left( {\left\{
{{n_\alpha }} \right\}} \right)}}} {\left. {\frac{{\partial \Xi
_{{n_1}, \ldots ,{n_p}}^k}}{{\partial {X_\beta }}}}
\right|_{{\textbf{X}^*}}},\nonumber \\\label{eq13}
\end{eqnarray}
where
\begin{eqnarray}
{\left. {\frac{{\partial \Xi _{{n_1}, \ldots ,{n_p}}^k}}{{\partial
{X_\beta }}}} \right|_{{\textbf{X}^*}}} =
\frac{{k!}}{{\prod\limits_\alpha {{n_\alpha }!} }}\left( {{n_\beta }
- {n_p}} \right){\left( {\frac{1}{p}} \right)^{k - 1}}.\label{eq14}
\end{eqnarray}
For $\alpha \neq \beta$, on the one hand, the contributions of the
state $\beta$ and the state $p$ to the summations in Eq.
(\ref{eq12}) and Eq. (\ref{eq13}) are equivalent with each other. On
the other hand, the summations contain the term $n_\beta-n_p$ in Eq.
(\ref{eq14}), such that the partial derivations of Eq.(\ref{eq12})
and Eq. (\ref{eq13}) are equal to zero. From Eq. (\ref{eq11}), we
conclude that all the non-diagonal elements of $\textbf{J}$ are
zero, i.e., $J_{\alpha \beta }=0$ for $\alpha \neq \beta$.
Furthermore, all the diagonal elements $J_{\alpha\alpha}$ of
$\textbf{J}$ are the same, $J_{\alpha\alpha}=J_{\beta\beta}$ for
each $\alpha$ and $\beta$, since all the states are symmetric.
Therefore, the eigenvalues of $\textbf{J}$ are $(p-1)$-fold
degenerate, given by $\Lambda(\textbf{J})=J_{\alpha\alpha}$. The
solution $\textbf{X}^*$ loses its stability whenever the eigenvalue
$\Lambda(\textbf{J})$ of $\textbf{J}$ is larger than 1, which yields
the critical noise,
\begin{eqnarray}
{f_{cB}} = \frac{{\sum\limits_k {\frac{{kP(k)}}{{\left\langle k
\right\rangle }}} {{\left. {\frac{{\partial P_k^\alpha }}{{\partial
{X_\beta }}}} \right|}_{{\textbf{X}^*}}} - 1}}{{\sum\limits_k
{\frac{{kP(k)}}{{\left\langle k \right\rangle }}} \left( {{{\left.
{\frac{{\partial P_k^\alpha }}{{\partial {X_\beta }}}}
\right|}_{{\textbf{X}^*}}} - {{\left. {\frac{{\partial \tilde
P_k^\alpha }}{{\partial {X_\beta }}}} \right|}_{{\textbf{X}^*}}}}
\right)}}.\label{eq15}
\end{eqnarray}
The other solutions $\textbf{X}\neq\textbf{X}^*$ ($m>0$) can be
obtained by solving Eq. (\ref{eq10}) numerically. Once $X_\alpha$
was found, one can immediately calculate $x_k^\alpha$ by Eq.
(\ref{eq5}) and ${m_\alpha } = \sqrt {p/(p - 1)} (\sum\nolimits_k
{P(k)x_k^\alpha }-1/p)$ by Eq. (\ref{eq1}).

\begin{figure}
\centerline{\includegraphics*[width=1.0\columnwidth]{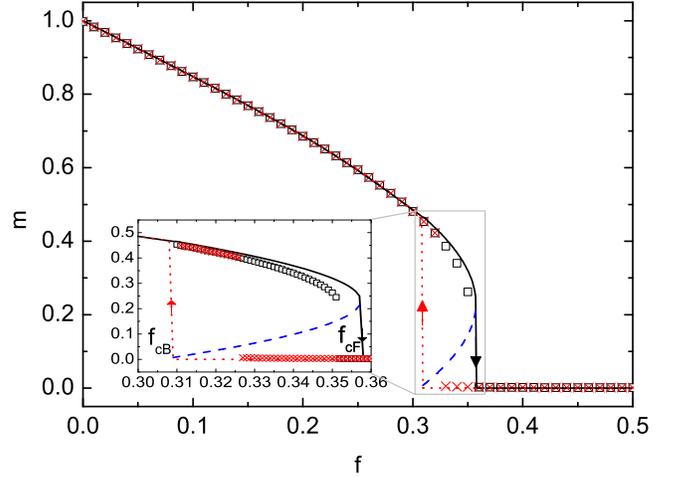}}
\caption{(Color online). Comparison between mean-field theory and MC
simulations for phase transition curve $m\sim f$ in the type-I
three-state MV model on ER networks. Lines correspond to the
theoretical results, and symbols to simulation ones. The networks
parameters used in the simulation are $N=5 \times 10^5$ and $\left
\langle k \right \rangle=10$. Within the hysteresis region, $m$ has
two stable solutions (black solid line and red dotted line) and one
unstable solution (blue dashed line). The inset shows an enlargement
for the hysteresis region. \label{fig3}}
\end{figure}

\begin{figure}
\centerline{\includegraphics*[width=1.0\columnwidth]{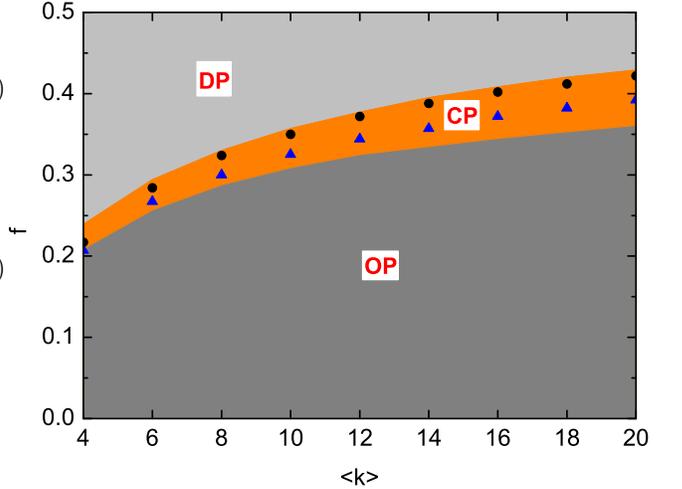}}
\caption{(Color online). Phase diagram in the type-I three-state MV
model on ER networks with different average degree $\left \langle k
\right \rangle$. The phase diagram is divided into three regions:
ordered phase (OP), disordered phase (DP), and coexisting phase
(CP). Symbols denote the simulation results on ER networks with $N=5
\times 10^5$: $f_{cF}$ (cycles) and $f_{cB}$ (triangles).
\label{fig4}}
\end{figure}

In Fig. \ref{fig3}, we show the theoretical results (lines) of the
type-I three-state MV model on ER random networks whose degree
distribution follows the Poisson distribution $P(k)={e^{ -
\left\langle k \right\rangle }}{\left\langle k \right\rangle^k}/k!$
with the average degree $\left\langle k \right\rangle =10$. The
theoretical calculation suggests that the type-I three-state MV
model undergoes a first-order order-disorder phase transition as $f$
varies. For $f<f_{cB}$, the only ordered phase with $m>0$ is stable.
For $f>f_{cF}$, the only disordered phase with $m=0$ is stable. In
the region $f_{cB}<f<f_{cF}$ (see the inset of Fig. \ref{fig3} for
an enlargement), two metastable phases with $m=0$ and $m>0$ coexist,
separated by an unstable state (dashed line). This leads to a
hysteresis phenomenon that is typical for a first-order phase
transition. For comparison, we also show the simulation results for
$N=5 \times 10^5$ in Fig. \ref{fig3}. There is excellent agreement
between our theory and the simulation outside of the hysteresis
region. However, a discrepancy exists between theory and simulation
for the prediction of phase transition points. One of the main
reasons may be that near phase transition points the lifetime of one
of the metastable states becomes short so that the metastable state
can not be fully sampled in the simulation. This is clearly realized
in Fig. \ref{fig1}: the simulation shows that $f_{cB}$ shifts to a
smaller value and $f_{cF}$ to a larger value as $N$ increases.

We consider the effect of the average degree $\left \langle k \right
\rangle$ on the phase transition in the type-I three-state MV model.
The results are summarized in Fig. \ref{fig4}. The phase diagram is
divided into three regions: the ordered phase (OP), the disordered
phase (DP), and the coexisting phase (CP) of OP and DP. With the
increase in $\left \langle k \right \rangle$, the coexisting region
is expanded and both the transition points shift to larger values.
The simulation results for $N=5 \times 10^5$ are also added into
Fig. \ref{fig4}, which agrees qualitatively with the theoretical
prediction.

We now demonstrate the nature of phase transitions for $p>3$. We
perform the theoretical calculation and MC simulation on ER networks
with $N=5 \times 10^5$ and $\left \langle k \right \rangle=10$ up to
$p=7$. For larger $p$, our theory is computationally prohibitive
since the high-dimensional summation in Eq.(\ref{eq8}) and
Eq.(\ref{eq9}) is time-consuming. The results show that for all
$p\geq3$ the phase transitions are of the first-order nature. The
phase transition points are shown in Table \ref{tab1}, from which
one can see that $f_{cB}$ is almost unaffected by $p$, and $f_{cF}$
increases monotonically with $p$ and approaches 0.5 as
$p\rightarrow\infty$.

\begin{table}[h]
\centering \caption{Phase transitions in the type-I $p$-state MV
model on ER networks. For $p \geq 3$, the phase transitions are
first order, essentially different from the second-order phase
transition in the two-state MV model. The simulation results are
obtained on networks with the size $N=5 \times 10^5$ and average
degree $\left \langle k \right \rangle=10$.} \label{tab1}
\begin{tabular*}{8cm}{@{\extracolsep{\fill}}llllll}
  \hline \hline
  \multirow{2}{*}{$p$} & \multirow{2}{*}{order} & \multicolumn{2}{c}{$f_{cB}$} &
\multicolumn{2}{c}{$f_{cF}$} \\ \cline{3-6}
  & & theo & simu   &  theo & simu \\ \hline

  2 & 2nd & 0.3091 & 0.296 & N/A & N/A \\
  3 & 1st & 0.3059 & 0.327 & 0.3573 & 0.350 \\
  4 & 1st & 0.3043 & 0.339 &0.4067 & 0.398 \\
  5 & 1st & 0.3038 & 0.350 &0.4429 & 0.434 \\
  6 & 1st & 0.3041 & 0.359 &0.4703 & 0.461 \\
  7 & 1st & 0.3055 & 0.360 &0.4918 & 0.483 \\
  \hline  \hline
\end{tabular*}\\
\end{table}

\begin{figure}
\centerline{\includegraphics*[width=1.0\columnwidth]{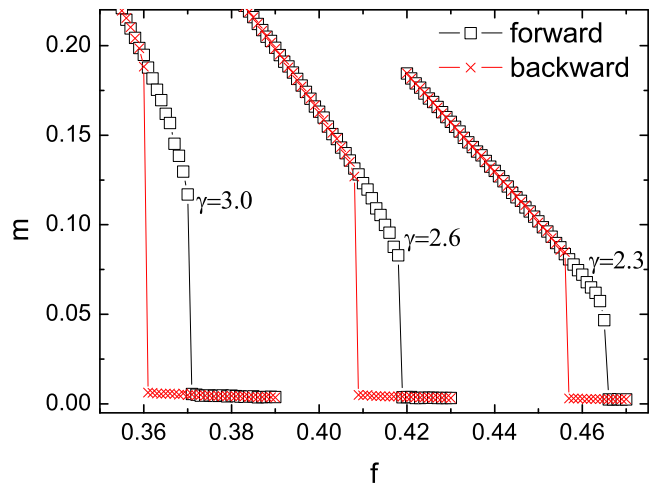}}
\caption{(Color online). Phase transition in the type-I three-state
MV model on scale-free networks with degree distribution $P(k) \sim
k^{-\gamma}$. From left to right, the degree distribution exponent
$\gamma=3.0$, 2.6, and 2.3. A larger $\gamma$ implies that the
network has higher degree heterogeneity. The results show that
degree heterogeneity suppresses the jump of magnetization near phase
transitions. The network size is $N=2 \times 10^5$ and the minimal
degree is $k_{min}=5$. Squares ($\Box$) and crosses ($\times$)
correspond to forward and backward simulations, respectively.
\label{fig5}}
\end{figure}

To consider the effect of degree heterogeneity on phase transition
in the type-I $p$-state MV model, we will show the results on
scale-free networks with degree distribution $P(k) \sim
k^{-\gamma}$.  The networks are generated by the configuration model
\cite{PRE.64.026118}.  Each node is first assigned a number of stubs
$k$ that are drawn from a given degree distribution. Pairs of
unlinked stubs are then randomly joined. This construction
eliminates the degree correlations between neighboring nodes.
Finally, we adopt an algorithm to reshuffle self-loops and parallel
edges that ensures the degree distribution is unchanged
\cite{PRE.70.06610}. In Fig. \ref{fig5}, we show $m$ as a function
of $f$ for several distinct $\gamma$. The larger $\gamma$ is, the
more heterogeneous the network is. The network size and the minimal
degree of nodes are fixed, $N=2 \times 10^5$ and $k_{min}=5$. One
can see that the nature of first-order phase transition does not
change with $\gamma$. As $\gamma$ increases, the jumps in $m$ at
phase transition points, $f_{cF}$ and $f_{cB}$ are depressed. We
have also considered some other $p$ and found that the main
conclusions are the same.

\subsection{Type-II dynamics}
\begin{figure}
\centerline{\includegraphics*[width=1.0\columnwidth]{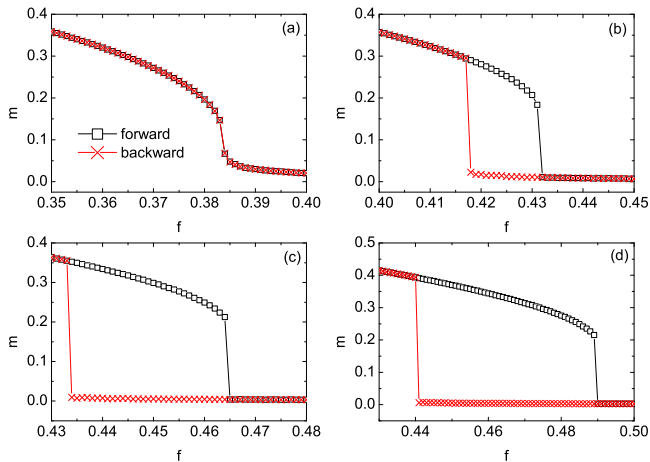}}
\caption{(Color online). Phase transition in the type-II $p$-state
MV model on ER networks. From (a) to (d), the numbers of states are
$p=3$, $4$, $5$, and $6$, respectively. For the type-II dynamics on
ER networks, the phase transition is first order for $p=3$ and
second order for $p\geq4$. The network size is fixed at $N=5 \times
10^5$. Squares ($\Box$) and crosses ($\times$) correspond to forward
and backward simulations, respectively. \label{fig6}}
\end{figure}

In this subsection, we consider the phase transitions in the type-II
$p$-state MV model. As shown in Fig. \ref{fig6}(a) for $p=3$, we
find that the forward and backward simulations coincide up to $N=5
\times 10^5$. This is a feature of continuous phase transition, in
agreement with \cite{JSM2010}, but in contrast with the result of
the type-I three-state MV model shown in Fig. \ref{fig1}. It is
interesting that such a small dynamical difference can lead to the
essential difference in the nature of phase transition in the
three-state MV model. For $p=4$, 5 and 6, as shown in Fig.
\ref{fig6}(b-d), we find that the phase transitions are
discontinuous, coinciding with the type-I dynamics.

\begin{figure}
\centerline{\includegraphics*[width=1.0\columnwidth]{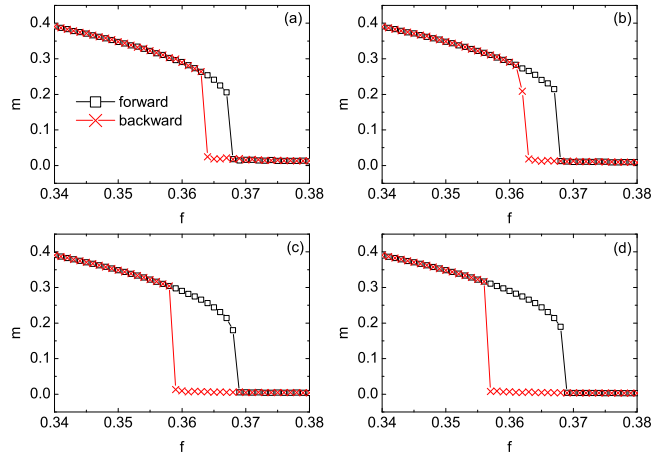}}
\caption{(Color online). First-order phase transition in the type-II
three-state MV model on degree-regular random networks, as opposed
to the second-order phase transition on ER networks shown in Fig.
\ref{fig6}(a). This shows that the order of phase transition in the
three-state MV model for the type-II dynamics depends on the
heterogeneity of degree distribution. From (a) to (d), the network
sizes are $N=5 \times 10^4$, $10^5$, $5 \times 10^5$, and $10^6$,
respectively. The degree of each node is exactly equal to $k=10$.
Squares ($\Box$) and crosses ($\times$) correspond to forward and
backward simulations, respectively. \label{fig7}}
\end{figure}

Moreover, as shown in Fig. \ref{fig5}, the degree heterogeneity can
suppress the discontinuity of magnetization at phase transition. A
natural question arises: Does a first-order phase transition happen
in more homogeneous networks than ER ones when the type-II dynamics
is taken into account? For this purpose, we show in Fig. \ref{fig7}
the three-state MV model on degree-regular random networks.
Interestingly, the phase transition now becomes first order. That
is, the nature of phase transition in the type-II three-state MV
model depends on the degree heterogeneity of the underlying
networks.

\section{Conclusions and Discussion}
In conclusion, we have studied numerically and theoretically the
order-disorder phase transitions in a $p$-state MV model on complex
networks. We find that for $p\geq3$ the phase transition is of a
first-order nature, significantly different from the second-order
phase transition in the original two-state MV model. A main feature
of the first-order phase transition is the occurrence of a
hysteresis loop as noise intensity goes forward and backward. Within
the hysteresis region, the ordered phase and disordered phase are
coexisting, and the rare phase flips can be observed due to the
finite-size fluctuation. The effects of the average degree and the
number $p$ of states on the two transition noises (i.e., the
boundaries of the hysteresis loop) are investigated. Also, we find
that degree heterogeneity can suppress the jump of magnetization at
phase transition. Moreover, we compare our model with that
introduced in \cite{PhysRevE.60.3666,JPA2002,JSM2010}. In spite of a
small difference in the dynamics, the types of phase transitions in
the three-state MV model on ER graphs are essentially different.
Interestingly, the phase transition for the latter dynamics becomes
first-order on degree-regular random networks. Therefore, the
dynamical rule and connectivity heterogeneity between agents play
important roles in the order of phase transitions in the three-state
MV model.

\begin{acknowledgments}
This work is supported by the National Natural Science Foundation of
China (Grants No. 61473001 and No. 11205002), the Key Scientific
Research Fund of Anhui Provincial Education Department (Grant No.
KJ2016A015) and ``211" Project of Anhui University (Grant No.
J01005106).
\end{acknowledgments}

%

\end{document}